National Research Council Canada

Conseil national de recherches Canada

Institute for Information Technology

Institut de Technologie de l'information

**NRC·CNRC**

# Design Guidelines for Landmarks to Support Navigation in Virtual Environments*

N.G. Vinson*Published in "Proceedings of CHI '99, Pittsburgh, PA. May 1999"Copyright 1999 by
National Research Council of Canada

Permission is granted to quote short excerpts and to reproduce figures and tables from this report, provided that the source of such material is fully acknowledged.Canada

NRC 43578

# Design Guidelines for Landmarks to Support Navigation in Virtual Environments

**Norman G. Vinson**
Institute for Information Technology
National Research Council, Canada
Ottawa, ON K1A 0R6
+1 613 993 2565
norm.vinson@iit.nrc.ca

**ABSTRACT**
Unfamiliar, large-scale virtual environments are difficult to navigate. This paper presents design guidelines to ease navigation in such virtual environments. The guidelines presented here focus on the design and placement of landmarks in virtual environments. Moreover, the guidelines are based primarily on the extensive empirical literature on navigation in the real world. A rationale for this approach is provided by the similarities between navigational behavior in real and virtual environments.

## INTRODUCTION

*Follow the road until you get to the church, then turn right. Then continue past two intersections. You'll see a gas station on one side of the road and a big apple tree on the other. Right after that, make your first left. At the stop sign, turn left again. I'll meet you in front of the house at the end of the road.*

Add the objects mentioned above to an environment you know well. Then imagine following the above directions in that context. Which objects stand out? The church, intersections, gas station, apple tree, stop sign, and the 'end of the road' probably stand out because they are reference points. Such distinctive environmental features functioning as reference points are **landmarks**. When associated to navigational actions (such as turn right), landmarks ease navigation by indicating when and where these actions should be taken.

Because of their navigational function, it is important to include landmarks in virtual environments (hereafter "**VEs**"). One can wonder though how to design and place landmarks in a VE to maximize their utility to the navigator. Research from the fields of urban planning (e.g. [19]), geography (e.g. [15]), and psychology (e.g. [3]) has explored the roles of landmarks in real world navigation. This empirical research can be applied to the development of design guidelines for VE landmarks to effectively support navigation. VEs designed according to these guidelines would facilitate users' navigation by permitting them to apply their real world navigational experience. The intent of this paper is therefore to provide landmark design guidelines to support navigation in large-scale VEs.

The guidelines herein focus on the structural elements and content of VEs because the virtual reality literature already contains many articles on specific navigational interfaces like input devices (e.g. [32]), motion control (e.g. [27]), and maps (e.g. [24]).

Before the guidelines themselves are presented, the necessity of supporting navigation in VEs is discussed, as is the justification for using research on real world navigation to create guidelines facilitating VE navigation.

### The Need for Navigational Support

The need for navigation design guidelines exists for three reasons: many VEs require the user to navigate, navigation in VEs is difficult, and disorientation is upsetting [18]. The first two reasons are discussed more fully below.

*VEs Often Require Navigation*
Navigation becomes necessary in environments that are so large that the navigator's viewpoint does not encompass the environment in its totality [3][7]. Such environments, whether virtual or real, are commonly termed "**large-scale environments**" (e.g. [3][26]). This scale forces the navigator to integrate the information provided by successive viewpoints into a coherent mental representation of the traversed environment, often termed "**cognitive map**" [3]. The navigator then relies on this cognitive map to navigate in the environment [25].

Current examples of large-scale VEs include simulators (flight, ship, or car) and some forms of telerobotics. As computers become more powerful and 3D visualization more common, fly-throughs of networks (electronic or biological) and complex molecular structures, simulations, and data visualizations will also pose navigational problems. The guidelines presented here are applicable to these types of large-scale VEs.

*VE Navigation Is Difficult*

Most of us have little difficulty navigating in the real world most of the time. The reason is that we mainly navigate in environments that are quite familiar. Even when environments are not completely familiar, we can often keep to familiar routes, for instance, by taking the same roads or buses. The difficulties arise when navigating unfamiliar environments. In these cases, we rely on navigational aids like written directions or maps. In urban environments, or on highways, we follow roads and signs that guide our travel. However, in unfamiliar environments these guides are often insufficient, and sometimes confusing [23]. Even maps can lead to navigational errors [35]. In natural environments, hikers use maps, and follow paths and signs. However, when a natural environment is devoid of such human artifacts, navigation is so challenging that it constitutes the competitive sport of orienteering [5].

The difficulty of navigating in unfamiliar real world spaces suggests the need to support navigation in VEs. A VE will always be unfamiliar when the user first encounters it. Gaining sufficient familiarity for successful navigation without any navigational support can take several hours that users may not be willing to provide [26]. Other differences between real and virtual environments increase the need to support navigation in VEs.

VEs contain fewer spatial and locomotive cues than real environments. Because of computational limitations, there is often less visual detail presented in VEs [26]. This means there may be fewer landmarks to support navigation, and fewer depth cues (such as occlusion and texture gradients (see [35])) to help with distance estimation. Locomotive and proprioceptive cues normally provided by walking, and turning one's body or head are often absent, especially from desktop virtual reality [26]. Finally, peripheral vision, absent from many forms of VE, has also been shown to provide navigational information [26]. These factors heighten the need for VE navigational support.

Moreover, the spatial structure of VEs may represent information. For example, a VE could contain objects whose spatial properties (e.g. shape, position, size) represent data values on different dimensions. Here, it is necessary for the navigator to quickly develop accurate representations of those spatial properties in order to understand the relationships in the data. In contrast, real environments typically do not represent data. Accordingly, in real environments, the most important function of navigators' spatial knowledge is to get them from here to there. This kind of navigation can be accomplished by remembering a string of associations between landmarks and their corresponding navigational actions (e.g. at the church, turn right) [17][29]. More accurate spatial knowledge is useful, for example to recover from navigational errors [35], but is not essential to navigation.

Thus, VEs that represent data demand greater accuracy of the navigator's cognitive map than do real environments. Guidelines promoting this accuracy are presented in the final section.

There is evidence that, despite these differences, the *way* in which we navigate is the same whether the environment is virtual or real. For instance, the development of spatial knowledge and its relation to navigation are the same for real and virtual environments [26]. Another example is that navigational experience with a virtualized environment has been found to transfer to the corresponding real environment [5]. Finally, principles and techniques coming out of real-world navigation research have been successfully applied to VE navigation research [1] [7]. This evidence provides a strong rationale for basing VE design guidelines on real-world navigation research.

In sum, it is clear that users of large-scale VEs require some navigational support. Moreover, it is reasonable to use research on real world navigation to generate guidelines for supporting navigation in VEs. The following section presents and explains such guidelines.

**GUIDELINES**

VEs should be easy to navigate, to leave cognitive resources available for the processing of any concurrent tasks. Because it is expected that people will navigate in unfamiliar VEs [7], VE design should promote rapid learning of the information necessary to navigate successfully. When information is represented by the relative size, orientation, or position of virtual objects, it is desirable that navigators develop accurate spatial information as quickly as possible.

These goals can be met by placing in the VE the types of objects that people use as cues for navigating in the real world: landmarks. Landmarks and their layout are critical for navigation [8][14][19]. In addition, the VE can be designed to be consistent with the way people remember large spaces. Cognitive maps are often distorted, but in predictable ways [3][35]. A designer who anticipates these distortions can minimize them by structuring the VE according to people's mnemonic predispositions.

In many cases, some features of the VE will be constrained by factors not under the designer's control. An example is provided by VEs with virtual objects that represent data. Designers have no control over such data objects, so the design guidelines cannot be applied to them. However, designers can add artificial landmarks to the environment, as long as they can be easily discriminated from the features and objects representing data. Those artificial landmarks can be designed and located according to the guidelines presented below. In this way, the designer can support navigation, while allowing the data objects to just represent data.

The first subsection contains explanations of how people use landmarks to learn the layout of an environment. This not only highlights the importance of landmarks, but also provides VE designers with a basic understanding of how cognitive maps are formed. In the following subsection, several abstract categories of landmarks and their role in navigation are presented. With this information, designers can review their VEs to ensure that all the landmark types are present. Guidelines on composing and placing landmarks to optimize their usefulness are then presented. These are followed by descriptions of environmental arrangements that minimize distortions in cognitive maps.

**Learning about an Environment**
Newcomers to an environment rely heavily on landmarks as points of reference [8][14]. As experience with the environment increases, navigators acquire *route knowledge* that allows them to navigate from one point in the environment to another [29]. Route knowledge is acquired and expanded by associating navigational actions to landmarks, such as turning right (action) at the corner (landmark). An ordered series of such action-landmark associations constitutes a route [12][17]. In following a route, the landmark (e.g. the corner) serves as the cue to recalling the associated action (e.g. turning right) [3].

In sum, landmarks support initial orientation in the new environment, they support the subsequent development of route knowledge, and they are essential to navigation using route knowledge. Thus, the first guideline is:

- **Guideline 1: It is essential that the VE contain several landmarks.**

Generally, additional experience with the environment increases the representational precision of route distances, and of the relative orientations and positions of landmarks [9][22]. Additional experience may also transform the representation from route knowledge to *survey knowledge* [29]. Survey knowledge is analogous to a map of the environment, except that it does not encode a typical map's top-down or bird's-eye-view perspective. Rather survey knowledge allows the navigator to adopt the most convenient perspective on the environment for a particular task [29][31]. Survey knowledge acquired through navigational experience also incorporates route knowledge [29].

In comparison to route knowledge, survey knowledge more precisely encodes the spatial proprieties of the environment and its objects [29]. Nonetheless, survey knowledge also contains distortions of the environment. Such distortions are especially problematic in VEs containing data objects (i.e. objects whose spatial properties represent data). The final subsection presents guidelines to minimize these distortions. Immediately below, the types of landmarks and their functions are discussed. The subsection following that contains guidlines on the construction and placement of landmarks.

**Landmark Types and Functions**
To include landmarks in a VE, one must know what constitutes a landmark. In his seminal work on urban planning and cognitive maps, Kevin Lynch found that people's cognitive maps generally contained five types of elements: paths, edges, districts, nodes, and landmarks. Each element *type* serves a particular function, though an *individual* element can serve more than one function[1] (see Table 1) [19].

Because these elements are used as landmarks (in the general sense[1]), and people make use of landmarks to navigate, the inclusion of Lynch's elements in a VE will support navigation through that VE. Moreover, since each type of element supports navigation in its own way, a VE designer should endeavor to include all five types of elements in the VE. Hence:

- **Guideline 2: Include all five types of landmarks (from Table 1) in your VE.**

**Table 1**: Landmark/Element Types and Functions.

| Types | Examples | Functions |
| --- | --- | --- |
| Paths | Street, canal, transit line | Channel for navigator movement |
| Edges | Fence, river | Indicate district limits |
| Districts | Neighborhood | Reference point |
| Nodes | Town square, public bldg. | Focal point for travel |
| Landmarks[1] | Statue | Reference point into which one does not enter |

**Landmark Composition**
It is important to include objects intended to serve as landmarks in a VE. However, it is also important that those objects be designed so that navigators will choose them as landmarks. There are two issues regarding the way in which landmarks should be constructed. One issue relates to the landmark's physical features. The other issue relates to the ways in which landmarks should be distinctive.

*Landmark Features*
A VE designer has the opportunity to create landmarks that are noticeable and help navigators remember their positions in the environment. Such landmarks support the use, and possibly the development of survey knowledge. For instance, a navigator can determine her position in the environment through her knowledge of the position of landmarks. Consequently, using particular features in designing landmarks can support navigation.

Evans and colleagues, expanding the work of Appleyard and Kaplan, empirically examined the relationship between building features and recall [11]. These studies produced a set of features that make a building more memorable, and a set of features that make the building's location easier to recall. Many of the features from both sets enhance a building's distinctiveness (see Table 2 and Guideline 3). Evans and colleagues found that the functions of buildings, their socio-cultural significance and their surrounding traffic patterns also affect their memorability. However, these types of features are more difficult to reproduce in a VE.

---

[1] Note that Lynch refers to these items as "elements" and reserves a specific meaning for the term "landmark" (see Table 1). In this paper, we use the term landmark more generally to refer to Lynch's elements and other features of an environment that provide information on navigator position and orientation.

- **Guideline 3: Make your landmarks distinctive with features from Tables 2 and 3.**

**Table 2**: Building Features Contributing to Memorability.

| Significant height [m] | Expensive building materials & good maintenance [l] |
|---|---|
| Complex shape [m] | Free standing (visible) [lm] |
| Bright exterior [l] | Surrounded by landscaping [m] |
| Large, visible signs [m] | Unique exterior color, texture [l] |

[m] Increases memorability of building.
[l] Improves memory for building location.

- **Guideline 4: Use concrete objects, not abstract ones, for landmarks.**

A study of VE landmarks also suggests that memorable landmarks increase navigability [26]. Landmarks consisting of familiar 3D objects, like a model car and a fork, made the VE easier to navigate. In contrast, landmarks consisting of colorful abstract paintings were of no help. It was felt that the 3D objects were easier to remember than the abstract art and that this accounted for the difference in navigability.

**Table 3**: Landmarks in Natural Environments.

| Manmade Items | Land Contours | Water Features |
|---|---|---|
| roads | hills | lakes |
| sheds | slopes | streams |
| fences | cliff faces | rivers |

While Lynch and Evans studied urban environments, Whitaker and colleagues examined the landmarks used in navigating natural environments. In a natural environment, any large manmade object stands out. Accordingly, experts in orienteering (natural environment navigation) relied most on manmade objects as cues when navigating [34]. They also used land contours and water features. However, they tried not to rely on vegetation since it is a rapidly changing, and therefore unreliable, feature in natural environments [33].

- **Guideline 5: Landmarks should be visible at all navigable scales.**

Finally, one can consider environment scales that differ from that of a city [12]. For instance, on a larger scale, a cognitive map of a country could have cities themselves as landmarks. It is not unusual for a user to have the ability to view a VE at different scales by "zooming in" or "zooming out". In such cases, it is important for the designer to provide landmarks at all the scales in which navigation takes place.

It is important to remember that the distinctiveness of an object is a crucial factor in its serving as a landmark. Consequently, it is important to apply the features from Tables 2 and 3 selectively. In the following section, this issue of distinctiveness is explored further.

*Distinctiveness*

- **Guideline 6: A landmark must be easy to distinguish from nearby objects and other landmarks.**
- **Guideline 7: The sides of a landmark must differ from each other.**

Objects intended to serve as landmarks must be distinctive in several ways. First, they must be distinctive in regard to nearby objects. Accordingly, Evans and colleagues note that a building that stands out from others on the same street is significantly more likely to be remembered [11]. Second, a landmark must be easy to distinguish from other landmarks, especially nearby ones. Otherwise, a navigator could confuse one landmark with another, and, as a result, select the wrong navigational action (e.g. make a wrong turn). This error is so common in the sport of orienteering, which involves navigation in natural environments, that it has been given a name: a parallel error [5]. Third, the sides of each landmark should differ from one another. These differences can help navigators determine their orientation. In contrast, consider a pine tree that is fairly symmetrical around the vertical axis. Because of its symmetry, the tree looks the same whether one is facing it from the East, from the West, or from any other direction in that plane. Consequently, navigators cannot use the tree to determine their orientation around the vertical axis (yaw). Without knowing one's own orientation, selecting a direction of travel to reach a destination becomes impossible. Navigation, other than aimless wandering, is therefore impossible. Accordingly, informal observation of navigators of a VE has revealed the superiority of asymmetrical landmarks in supporting navigation [6].

- **Guideline 8: Landmark distinctiveness can be increased by placing other objects nearby.**

Where a single object is not distinctive enough, a pair of objects may suffice. Consider again the radially symmetrical pine tree. Due to the tree's symmetry around the vertical axis, it is difficult to tell from which direction one is viewing it. However, by inserting a lamppost next to the tree, the viewing direction is disambiguated. From one direction, the lamppost is in front of the tree. From the opposite direction, the tree hides the lamppost. Viewed from another direction the tree is to the *left* of the lamppost, while viewed from the opposite direction the tree is to the

*right* of the lamppost. This technique can also differentiate the views from directions falling around the horizontal axes. Moreover, it can be used to differentiate two similar landmarks. Here, one need only insert a different object near one (or each) landmark. The designer should keep in mind however that it is likely to be more difficult for navigators to determine their positions or orientations under these circumstances. With several objects, the spatial relationships between them must be considered in order to determine the viewpoint's orientation or position. This processing is not required when a single object provides unambiguous position or orientation information. It is the additional processing required with several objects that probably makes it more difficult for navigators to estimate their positions and orientations from several objects. Nevertheless, some VE navigators have been observed relying on configurations of landmarks to obtain orientation information [6].

- **Guideline 9: Landmarks must carry a common element to distinguish them, as a group, from data objects.**

Finally, consider VEs whose features are constrained by the underlying data, such as the human circulatory system. Although some of the objects in these VEs can serve as landmarks, it is possible to further assist navigation by augmenting the VE with additional objects that only function as landmarks. However, navigators must easily recognize these objects as landmarks and realize that they are *only* landmarks. To continue the circulatory example, navigators should not take an artificial landmark for a blood cell. Otherwise, navigators could develop misconceptions about blood cells (e.g. about their motion, position, or shape). Consequently, in such VEs it is important for landmarks to carry some common element that distinguishes them from the other virtual objects. For instance, artificial landmarks in the circulatory system could appear solid and angular, like a truck, in contrast to the soft, bulbous objects shown travelling through the blood. Nonetheless, these artificial landmarks must still be distinctive as described above.

In sum, it is possible to compose VE landmarks with the features navigators use to select landmarks in the real world. Thus, navigators' experience and navigational abilities can transfer from real environments to virtual ones. The common theme in the selection of landmarks seems to be their distinctiveness. However, not only the appearance of landmarks is important to navigation. Their placement must also be carefully considered, as we see below.

**Combining Paths and Landmarks: Landmark Placement**
- **Guideline 10: Place landmarks on major paths and at path junctions.**

Evans and colleagues found that the memorability of a building and its position was also affected by the building's location in the environment [11] (Table 4). In short, memorability is enhanced when the building is located on a major path or at a path junction (see also [14]). These findings highlight the importance of including paths in a VE to provide a structure for placing landmarks. The correct placement of landmarks enhances their memorability and consequently, eases navigation.

**Table 4**: Building Positions Contributing to Memorability.

| Located on major path [m] | Visible from major road [lm] |
|---|---|
| Direct access from street (esp. no plaza or porch) [lm] ||
| Located at important choice points in circulation pattern [m] ||

[m] Increases memorability of building.
[l] Improves memory for building location.

Paths can also facilitate navigation by guiding the navigator to points of possible interest. Moreover, paths provide a way for the designer to minimize the number of landmarks in the VE while still supporting navigation. The recommendation to use to landmarks to support navigation can be problematic for a designer of large-scale VEs. To maintain an acceptable level of interactivity, a VE designer will want to limit the number and complexity of virtual objects in the environment. On the other hand, to support navigation, especially the acquisition and use of route knowledge, the designer must include many landmarks in the VE. Specifically, when at least two landmarks can be seen from each viewpoint, navigators can represent a route from one place to another as a string of landmarks. This allows navigators to follow the whole route by moving from one landmark to the next. Supporting all possible routes through a VE in this way would require a vast number of landmarks. Paths allow the designer to minimize the number of landmarks, thus enhancing interactivity, while still supporting navigation. Here, many of the landmarks are placed at path junctions, and a few others are placed along the paths themselves. Navigators use the landmarks along the paths for distance estimation and course verification [33][34]. When navigators reach a path junction, they select the appropriate navigational action (e.g. turn left) via landmark recognition. Thus, paths support the acquisition and use of route knowledge with fewer landmarks.

In sum, it is important to use both paths and landmarks to support navigation; especially navigation based on route knowledge. Nonetheless, when the spatial properties of virtual objects represent data, route knowledge is insufficient to provide an understanding of the relationships in the data. Survey knowledge is needed for this purpose. The common distortions in survey knowledge must also be minimized. Guidelines for this are presented in the following section.

**Minimizing Distortions in Cognitive Maps**
The term "cognitive map" is misleading in that it suggests that mental representations of environments are very much like images. In reality, our cognitive maps contain many features that are not image-like. Cognitive maps contain categorical and hierarchical structures [3][16][31], and many spatial distortions, some of which cannot even be represented in an image [15][20][30][35]. Distance asymmetry is an example of such a distortion. A distance asymmetry involves representing two different distances between two points, e.g. A and B, wherein the A to B distance is different from the B to A distance. Both of these distances cannot be represented in a single image. Since people use their cognitive maps to navigate [25], distortions in their cognitive maps can be confusing and lead to navigational errors. Moreover, such distortions will warp a navigator's understanding of data represented by data objects. For these reasons, it is important to minimize these distortions. Studies have shown that these distortions diminish with increased navigational experience [9][13][22]. However, it is possible to structure an environment to minimize the development of these distortions in the first place. This requires an understanding of the types of distortions and their causes. Accordingly, this is presented first, followed by an examination of the design guidelines.

*Sources of Distortions*

Many distortions seem to result from the hierarchical structure of cognitive maps [3][16]. These hierarchies can be formed by clustering objects that fall within identifiable boundaries to form districts (see Table 1). For instance, all the cities in one state (a district) can form a cluster. When there are no objective boundaries, clusters can form around anchor points, which can be important landmarks [4][6][16]. A multi-level hierarchy can develop wherein districts themselves are clustered at a higher level [16]. An example of a hierarchy-induced distortion is provided by Stevens and Coupe [28]. Participants in one experiment reported that San Diego California is west of Reno Nevada. This response resulted from an inference based on the hierarchical relationship between the cities and their containing states. People reasoned that since San Diego is in California, Reno is in Nevada, and that California is west of Nevada, then San Diego must also be west of Reno. Distances can also be distorted. Here, people underestimate the distances between objects in the same district, while overestimating distances between objects from different districts [16]. In sum, cognitive map hierarchies can produce distortions of relative directions and distances.

Distortions are also produced by mental heuristics that help us remember the layout of objects [30]. One heuristic aligns the main axes of nearby objects. Main axes are provided by an object's shape and/or its most salient features. For example, people drew a map of Palo Alto California showing familiar streets as being more parallel than they are in reality. Another heuristic is to rotate an object relative to its background, so the object's main axes line up with the background's [30]. For example, residents of Umeå Sweden misrepresented the northern direction so that it corresponded with Umeå's street grid [13]. Tversky speculates that an environment's axes (e.g. the axes of Umeå's street grid) provide a possible framework on which to construct the type of hierarchy described above. She also notes that the alignment and rotation heuristics could be responsible for rectilinear normalization, the tendency to distort environmental features into a grid [30][35].

In sum, hierarchical structures and Tversky's heuristics can explain several, though not all, of the distortions found in cognitive maps. By understanding how distortions come about, we can design VEs to minimize them. Three guidelines for doing so are discussed in the following section.

*The Grid*

- **Guideline 11: Arrange paths and edges (see Table 1) to form a grid.**
- **Guideline 12: Align the landmarks' main axes with the path/edge grid's main axes.**
- **Guideline 13: Align each landmark's main axes with those of the other landmarks.**

To minimize the distortions, the designer must create a VE that induces a hierarchical representation whose districts form a grid. A consequence of the grid's spatial regularity is that the spatial relationships between districts are a good approximation of the spatial relationships between objects in those districts. For instance, if district A is to the left of district B, then all objects in A are to the left of all objects in B. Consequently, judgments about the relative positions of objects are not so distorted, even though these judgments are based on the spatial relationships between districts. Figure 1 shows theoretical direction distortions for districts forming a grid and an irregular shape. (The irregular districts correspond to the California and Nevada example from the Sevens and Coupe study discussed previously.) A grid structure still produces some distortions, but they are smaller than those produced by an irregular structure.

**Figure 1:** Cognitive Map Direction Distortions in Grid-Form and Irregular Districts.

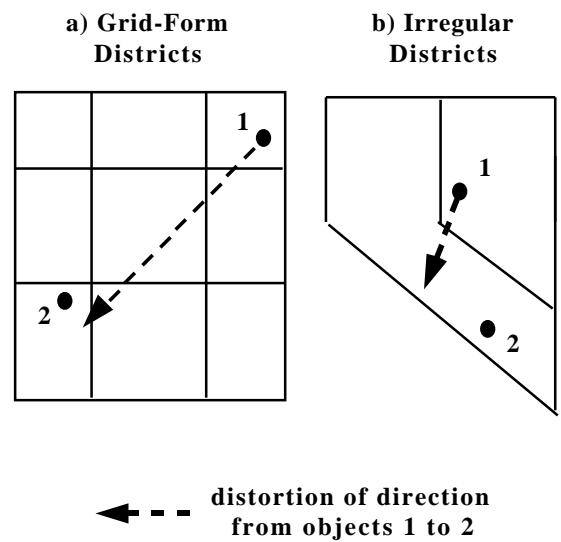

Accordingly, it has been shown that distance and direction judgements are more accurate in environments with street grids [10][13]. The question then is how to encourage the navigator to construct a grid-like representation? It appears that people use cues in an environment to structure their spatial hierarchies [6][16][30]. Consequently, the designer should arrange the environment's paths and edges to form a grid. Each landmark's main axes should be aligned with the other landmarks' main axes and the path/edge grid's too. This will reinforce the grid-like districting. Moreover, it will substantially reduce distortions due to rotation and alignment. Since the landmarks are already aligned, the navigator will not need to bring them into alignment by distorting their position. Since the landmarks' axes and the environment's axes already coincide, the navigator will not need to rotate the landmarks or the environment. Finally, rectilinear normalization does not need to be performed since the objects in the environment are already rectilinear.

Darken and Sibert placed a generic grid on the surfaces of their VEs. However, the grid's structure was incompatible with the environments' (i.e. the above guidelines were violated). The grid did improve navigability, but statistical comparisons between the grid and no grid conditions were not computed. Some analyses revealed that the grid could *interfere* with the acquisition of survey knowledge [7]. This most likely occurred because of the structural incompatibility between the grid and VEs. These results reveal the importance of following the guidelines presented above.

## CONCLUSION

This paper focused on the use of landmarks in human navigation. Landmarks not only indicate position and orientation, but also contribute to the development of spatial knowledge. Therefore, a VE containing distinctive landmarks supports navigation by facilitating the acquisition and application of spatial knowledge.

The substantial research on human navigation in real environments was used to formulate guidelines for landmark design. Guidelines to increase the accuracy of a navigator's spatial knowledge were also presented. Because the guidelines are based on real-world navigation, VE navigators are encouraged to transfer their real-world navigational abilities. Consequently, following these guidelines in constructing VEs will make them more navigable.

The guidelines presented here can be considered design rules-of-thumb – untested generalizations from one domain to another. Brooks notes that VE designers are in need of such rules-of-thumb [2]. Accordingly, these guidelines can be of use to VE designers who have little research interest in navigation. Specifically, navigational problems can interfere with concurrent tasks that are the topics of research. The VE designer can follow these guidelines to ease navigation and thus allow users to focus on the tasks of interest. On the other hand, where navigation is the research topic, these guidelines can be used as starting points for empirical testing or further hypothesis generation.

## ACKNOWLEDGEMENTS
The author thanks Janice Singer and Marceli Wein for their comments.